\documentclass[a4paper,11pt]{article}
\usepackage{fullpage}
\usepackage{times}
\usepackage{subfigure}
\usepackage{vmargin}
\usepackage{fancyheadings}
\usepackage[cp1251]{inputenc}
\usepackage[english]{babel}
\usepackage{graphicx}
\usepackage{color}
\usepackage{cite}
\usepackage{tabularx}
\usepackage{amsmath,amssymb,amsthm,enumerate,bbm,color,verbatim,setspace}

\newcommand{\sign}{\text{sign}}
\usepackage{braket}
\theoremstyle{plain}

\theoremstyle{remark}

\date{}
\title{HIGH-PRECISION OBSERVABLE ESTIMATION 
WITH SINGLE QUBIT QUANTUM MEMORY}

\author{L.A. Markovich$^{1,2,3,4*}$, J. Borregaard$^{2}$\\
      $^1$Instituut-Lorentz, Universiteit Leiden, P.O. Box 9506,\\ 2300 RA Leiden, The Netherlands\\
 $^2$QuTech and Kavli Institute of Nanoscience, Delft University of Technology,\\ 2628 CJ, Delft, The Netherlands\\
$^3$Institute for information transmission problems, Moscow,\\ Bolshoy Karetny per. 19, build. 1, Moscow 127051, Russia\\
$^4$Russian Quantum Center, Skolkovo, Moscow 143025, Russia\\
$^*$Corresponding author e-mail: markovich@mail.lorentz.leidenuniv.nl}

\date{}
\begin{document}
\maketitle
\pagenumbering{arabic}
\begin{abstract}\noindent The estimation of multi-qubit observables is a key task in quantum information science. The standard approach is to decompose a multi-qubit observable into a weighted sum of Pauli strings. The  observable can then be estimated from projective single qubit measurements according to the Pauli strings followed by a classical summation. As the number of Pauli strings in the decomposition increases, shot-noise drastically builds up, and the accuracy of such estimation can be considerably compromised. 
Access to a single qubit quantum memory, where measurement data may be stored and accumulated can circumvent the build-up of shot noise. Here, we describe a many-qubit observable estimation approach to achieve this with a much lower number of interactions between the multi-qubit device and the single qubit memory compared to previous approaches. 
Our algorithm offers a reduction in the required number of measurements for a given target variance that scales $N^{\frac{2}{3}}$ with the number of Pauli strings $N$ in the observable decomposition. The low number of interactions between the multi-qubit device and the memory is desirable for noisy intermediate-scale quantum devices.\end{abstract}
\medskip

\noindent{\bf observable estimation; quantum decoherence; single qubit quantum memory; multi-qubit states.} 


\setcounter{section}{-1} 
\section{Introduction}
\par Determining the expectation value of multi-qubit observables within a quantum system is a fundamental subject in numerous domains of quantum science. Notably, in condensed matter physics, materials science, quantum chemistry, and combinatorial optimization~\cite{Golub2000}, the objective revolves around identifying spectral characteristics, such as the ground state energy or the lowest eigenvalue of a Hamiltonian. However, the direct estimation of the observable expectation value is not a straightforward task in these scenarios.
 The quantum phase estimation (QPE) algorithm~\cite{Kitaev1995,Kimmel2015,Berg2020,mohammadbagherpoor2019,Wiebe2016,O_Brien2019}, offers a potential solution for ideal quantum processing units with extended coherence times. However, in the Noisy Intermediate-Scale Quantum (NISQ) era, QPE's implementation is hindered by coherence time limitations. To overcome this, the quantum energy (expectation) estimation (QEE) method is widely used, especially within the variational quantum eigensolver framework~\cite{Peruzzo2014}.
\par QEE requires just single qubit measurements and thus a minimum coherence time of the quantum device. However, it is not without its limitations. One such drawback is the accumulation of shot noise during the estimation process, which leads to a reduction in the overall variance level. In QEE, the individual Pauli strings are estimated separately, and subsequently, a linear combination of these estimates is utilized to determine the value of the observable.
\par Consequently, in order to estimate an observable comprised of $N$ Pauli strings with a variance of $\eta$, each Pauli string should be estimated with an variance of $O(\eta/N)$. The resulting sample complexity therefore scales as $O(N^2)$. This can pose a significant challenge since the total number of measurements will eventually be limited by the available run-time of the quantum device. It is important to note that the measurement process itself is often one of the most time-consuming operations in current quantum devices~\cite{jurcevic2021,brown2016,schafer2018}.
\par Recent studies have proposed strategies to minimize sample complexity by aggregating commuting sets of Pauli strings~\cite{Wang2019,Hamamura2020,Crawford2021}, serving as alternatives between QPE and QEE. While these approaches improve efficiency and reduce quantum resource requirements for obtaining the observable's expectation value, they face challenges. Notably, they do not address the fundamental scaling issue related to noise accumulation with the increasing number of Pauli strings in the observable decomposition~\cite{McClean2016,Crawford2021}.
\par In our recent paper \cite{https://doi.org/10.48550/arxiv.2212.07710}, we introduced a novel  algorithm known as the Coherent Pauli Summation (CPS) method, which effectively mitigates the issue of shot-noise accumulation by leveraging a single-qubit quantum memory (QM). The pivotal aspect of the CPS method lies in the utilization of Quantum Signal Processing (QSP) techniques~\cite{PhysRevLett.118.010501}, to encode the mean value of Pauli strings within the phase of a single qubit QM. By circumventing the accumulation of shot noise, the CPS method achieves a significant improvement in the variance of the estimate compared to the Quantum Energy Estimation (QEE) method, scaling at $O(N)$, where $N$  is the number of Pauli strings in the observable decomposition. However, the QSP step necessitates multiple controlled many-qubit unitaries between the single qubit quantum memory and the many-qubit NISQ device for the encoding of each Pauli string. This can have a detrimental effect on the coherence of the quantum memory given the complexity of such operations. In general, it is desirable to limit the number of interactions with the quantum memory as much as possible to uphold the coherence of the stored information.  
\par A single-qubit QM typically consists of a physical system, such as an individual ion~\cite{langer2005long,wang2017single} or Rydberg atom~\cite{li2016quantum}, capable of storing and maintaining the quantum state of a qubit. Such QM performs well while isolated from the environment to have low decoherence rates and long coherence times to preserve information.
Another possibility is an error-corrected QM~\cite{RevModPhys.87.307,doi:10.1080/09500340.2012.737937}. Its basic idea is to use a larger number of physical qubits to represent each logical qubit. These physical qubits are entangled with each other and form an error-correcting code, which allows the system to detect and correct errors without losing the encoded quantum information. Even if some of the physical qubits experience errors, the encoded logical qubit can be recovered with high fidelity. While error-corrected QM holds significant promise in mitigating errors and enhancing the reliability of quantum information processing, it also comes with several  disadvantages like increasing complexity and reducing the processing speed. Moreover, the process of error correction itself introduces a possibility of logical errors due to imperfect gate operations or residual errors in the encoding.  
The CPS requires $N\log{(N/\sqrt{\eta})}/\log{(\log{(N/\sqrt{\eta})})}$ controlled unitary operations  between the NISQ qubits and the memory qubit that is a challenging demand for the current QMs.
\par In this article, we present an alternative approach to our CPS method. Instead of relying on QSP techniques for encoding the mean value of Pauli strings in the phase, we propose the utilization of a Taylor series approach. By employing the Taylor-based CPS method (TCPS), we achieve a variance improvement of the final estimate compared to the QEE method, scaling as $N^{2/3}$. While the coherence time of the memory qubit required for both TCPS and the original CPS method scales linearly with $N$, the TCPS method only requires a single controlled unitary between the memory qubit and the multi-qubit device for the encoding of a Pauli string compared to the $O(\log(1/\eta))$ number of unitaries required for the CPS method, where $\eta$ is the target precision of final multi-qubit observable estimate. Although the TCPS approach exhibits a slight scaling disadvantage compared to the original CPS method, it offers the advantage of fewer interactions between the single qubit memory and the multi-qubit device which can important for practical implementations.
\par The paper is organised as follows.
In Section~\ref{sec:1} we briefly outline the QEE method.
In Section~\ref{sec:2} the TCPS method is studied in details. We compare it with the QEE method in terms of variance of the resulting estimate and the resources. All technical details are given in Appendix.

\section{QEE Technique}\label{sec:1}
\par We start by going through the main steps of the  QEE technique.
Decomposition of an observable  $O$ into a weighted sum of Pauli strings 
\begin{eqnarray} \label{eq:0}
    O=\sum\limits_{j=1}^N a_{j}P_{j},\quad  a_j\in \mathrm{R}
\end{eqnarray}
is a crucial step in the QEE for determining the expected value of a given observable for a particular quantum state $\ket{\Psi}$. 
Here  $a_j\in R$ are the  decomposition coefficients and $P_{j}$ are the Pauli strings composed as tensor products of single qubit Pauli matrices and the identity. 
Since a set of $d^2$ Pauli strings form a complete operator basis for a Hilbert space with dimension $d$ this decomposition is always possible. 
\par
\begin{figure}[ht!]  
\centering 
\includegraphics[width=0.8\linewidth]{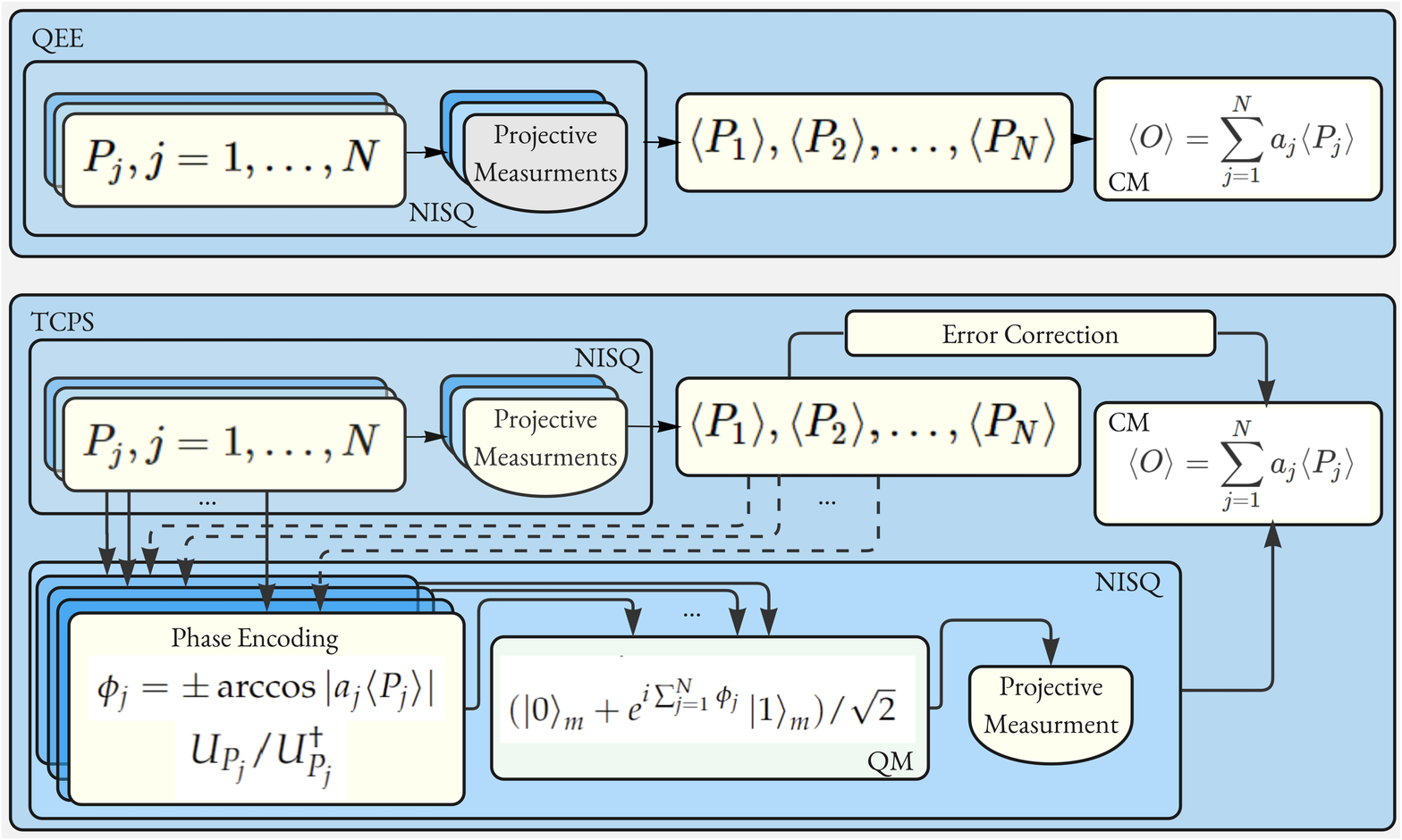} 
\caption{The high-level picture of QEE and TCPS methods. In QEE the expectation value of each Pauli string is estimated by a series of projective measurements. Then  all $\langle \hat{P}_j\rangle$, $j=1,\dots, N$ are summed   to obtain an estimate of the observable $\langle O\rangle$. In the TCPS method all $\langle \hat{P}_j\rangle$ are encoded in the single qubit QM by a sequential phase kick-back algorithm. Also a small amount of projective measurements of each Pauli string is done to estimate the sign of the Pauli strings and classical correction of the final estimate. After encoding of all Pauli strings, a projective measurement on the QM qubit is performed and the whole process is repeated to obtain an estimate of $\langle O\rangle$ to the variance $\eta$.
}\label{fig:4}   
\end{figure}

 The Pauli strings are measured sequentially or in parallel  (if the strings are commuting), using single qubit projective measurements in order to provide an estimate of each Pauli string  $\langle P_j\rangle\equiv\langle \Psi|P_j|\Psi\rangle$~\cite{Peruzzo2014,McClean2014}.
The mean value of the observable $O$ is than calculated by the classical summation of the later estimates (see Fig.~\ref{fig:4}):
\begin{eqnarray}\label{eq:1}
    \langle O\rangle=\sum\limits_{j=1}^{N}a_j\langle P_j\rangle.
\end{eqnarray}
It follows that the shot noise from the individually estimated mean values of the Pauli strings accumulates in the final estimate of $\langle O\rangle$. 
If every  $\langle P_j\rangle$ is estimated with a variance  $\sigma^2(\langle\hat{P}_j\rangle)$ then, assuming equal weights of the Pauli strings in \eqref{eq:0}, the variance of estimate of $O$ is  $\sigma^2(\hat{\langle O\rangle})\sim N\sigma^2(\langle\hat{P}_j\rangle)$. 
\par Since the dimension of the Hilbert space increases exponentially with the number of qubits, the number of Pauli stings $N$ in the decomposition \eqref{eq:0}  can be very large. In particular, the encoding of fermionic Hamiltonians in a qubit lattice  poses a problem for quantum simulators of fermionic systems. In \cite{jordan1993paulische}, local fermionic Hamiltonian systems are mapped to local spin Hamiltonians (fermion-to-qubit mapping) suitable for analog quantum simulation applications. However, such transformations introduce long-range many-body interaction terms, which are beyond the scope of the current implementation and necessitate special reduction algorithms \cite{https://doi.org/10.48550/arxiv.1706.03637}.
\par To overcome the accumulation of shot noise we  employ \textit{phase-kick back} techniques to encode the mean value of each Pauli string into the phase of a single qubit QM, allowing the encoding of $\langle O\rangle$ which can be directly measured.

\section{Main Results}\label{sec:2}
\par Let  $|\Psi_0\rangle \equiv V|\pmb{0}\rangle$ be a quantum state of the NISQ device, where $V$ is an invertible preparation circuit. By $|\pmb{0}\rangle$ we denote the state where all qubits are prepared in their ground states $\ket{0}$.
Our target is to estimate the expectation value $ \langle \Psi_0| O|\Psi_0\rangle$ with a variance of $\eta$. 
To this end, we define
$|\Psi_1\rangle_j\equiv P_j|\Psi_0\rangle$  and 
introduce   the unitary  operator 
\begin{eqnarray}\label{eq:3}
  U_{P_j}=V\Pi_0V^{\dagger}P_j,  
\end{eqnarray}
where  $\Pi_0=I-2\ket{\pmb{0}}\bra{ \pmb{0}}$ is a projector, and $I$ is the identity operator.
This operator defines a rotation by a principal angle \begin{eqnarray}\label{eq:4}\phi_j=\arccos{|\langle\Psi_0|\Psi_j\rangle|}\end{eqnarray} between two  closed subspaces $|\Psi_0\rangle$ and $|\Psi_j\rangle$ of a Hilbert space. It holds that the
state $|\Psi_0\rangle$ is an equal superposition of eigenstates $|\phi^{\pm}\rangle$  of $U_{P_j}$ with eigenvalues $e^{\pm i\phi}$, respectively~ \cite{Knill2007,Wang2019}.  
\par We introduce the state
\begin{eqnarray}
    \left(\sqrt{(1-\epsilon'_j)}|0\rangle_p+\sqrt{\epsilon'_j}|1\rangle_p\right)\otimes|\Psi_0\rangle,\quad  \epsilon'_j\in(0,1),
\end{eqnarray}
where the first register describes the processing ancillary. Acting on this state with a unitary operator $|0\rangle_p\langle 0|\otimes I+|1\rangle_p\langle 1|\otimes cP_j$, we get the following $\epsilon'$-dependent state
\begin{equation}
  |\tilde{\Psi}_0\rangle_j\equiv \sqrt{1-\epsilon'_j}|0\rangle_p|\Psi_0\rangle+\sqrt{\epsilon'_j}|1\rangle_p|\Psi_j\rangle.
    \label{eq:3_1}
\end{equation}
Let us introduce $|\tilde{\Psi}_1\rangle\equiv (\sigma_x\otimes I)|\tilde{\Psi}_0\rangle$ and calculate
the overlap \begin{eqnarray}\label{eq:6}
|\langle \tilde{\Psi}_0|\tilde{\Psi}_1\rangle_j| =2\sqrt{\epsilon'(1-\epsilon')}|\langle \Psi_0|\Psi_j\rangle|.
\end{eqnarray}
  To encode the weighted sum of the Pauli strings we  select  $\sqrt{\epsilon'_j(1-\epsilon'_j)}\equiv |a_j| \sqrt{\epsilon}$, where   the parameter $\epsilon\in(0,1)$, holds.
\par The rotation operators $(\tilde{U})_{P_j}(\epsilon)=\tilde{V}\Pi_0\tilde{V}^{\dagger}P_j$, $\tilde{V}(P_j):\tilde{V}(P_j)|0\rangle=|\tilde{\Psi}_0\rangle_j$ can be introduced to encode the Pauli strings in the phases. 
The operator $(\tilde{U})_{P_j}(\epsilon)$ defines the rotation by a principal angle \begin{eqnarray}\label{eq:4}\tilde{\phi}_j=\arccos{(2\sqrt{\epsilon}|a_j\langle P_j\rangle|)}\end{eqnarray} between two  closed subspaces $|\tilde{\Psi}_0\rangle$ and $|\tilde{\Psi}_1\rangle$ of a Hilbert space.
Since this encoding method requires preparation of the eigenstates $|\phi^{\pm}\rangle$ of the rotation operators at each
iteration,  it is important to mention, that we can write $|\tilde{\Psi}_0\rangle$  as a superposition 
\begin{eqnarray}\label{eq:7}
  &&  |\tilde{\Psi}_0\rangle
=\frac{1}{\sqrt{2}}(|\tilde{\phi}_0^+\rangle+|\tilde{\phi}_0^-\rangle), \\\nonumber
&&|\tilde{\phi}_0^+\rangle\equiv (\sqrt{1-\epsilon'}|0\rangle+\sqrt{\epsilon'}|1\rangle)|\phi^{+}\rangle,\quad |\tilde{\phi}_0^-\rangle\equiv (\sqrt{1-\epsilon'}|0\rangle+\sqrt{\epsilon'}e^{i\phi}|1\rangle)|\phi^{-}\rangle.
\end{eqnarray}
Hence, projecting on $|\tilde{\phi}_{0}^{\pm}\rangle$ states is equivalent to projecting on $|\phi^{\pm}\rangle$, respectively. 
\par To encode the information about the mean value of the Pauli string in the phase we define a memory qubit in the state $\ket{\psi}_m=\ket{+}_m$. The controlled version  c$(\tilde{U})_{P_j}(\epsilon)$ can be implemented by substituting $P_j$ with its controlled version, c$P_j$. Applying c$(\tilde{U})_{P_j}(\epsilon)$ to the state $\ket{\psi}_m|\tilde{\Psi}_0\rangle$, we get 
\begin{equation}
   cU_{P_j}\ket{\psi}_m|\Psi_0\rangle\longrightarrow \frac{1}{\sqrt{2}}\left(\frac{|0\rangle_m+e^{i\phi_j}|1\rangle_m}{\sqrt{2}}|\phi_j^{+}\rangle+\frac{|0\rangle_m+e^{-i\phi_j}|1\rangle_m}{\sqrt{2}}|\phi_j^{-}\rangle\right),
    \label{eq:5}
\end{equation}
with the information about the mean value of the Pauli string encoded in the phase, i.e.
\begin{eqnarray}
   |\langle\Psi_0|P_j|\Psi_0\rangle|=\cos{(\phi_j)}. 
\end{eqnarray} 
\par
\begin{figure}[ht!]  
\centering 
\includegraphics[width=0.8\linewidth]{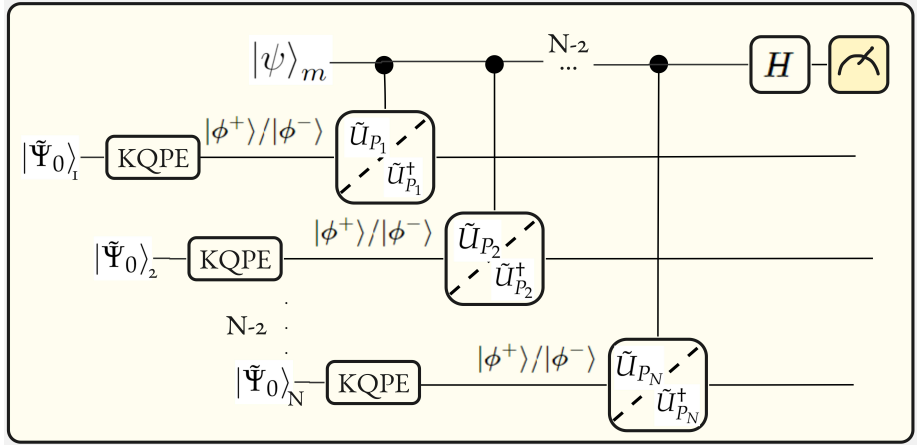} 
\caption{The  phase encoding  in the QM is shown. The QM qubit is prepared in $\ket{\psi}_m=\ket{+}_m$ state. We prepare every $\ket{\tilde{\Psi}_0}_j$ state and do $n_{QPE}$ steps of KQPE to resolve between 
$\ket{\phi^+}$ and $\ket{\phi^-}$  eigenstates. After we act on the state with $\tilde{U}_{P_j}$ or $\tilde{U}_{P_j}^{\dagger}$, respectively to encode the phase $\tilde{\phi}_j$ into the QM. We address the QM $N$ times to do the whole round of encoding. Finally the projective measurement is performed. 
}\label{fig:5}   
\end{figure}

 Using this encoding method, we can encode the correct linear combination of the $N$ mean values of Pauli strings sequentially in the phase of the memory qubit as depicted in Fig.~\ref{fig:5}.
The final state of the memory qubit will be $(|0\rangle+e^{i\tilde{\Phi}}|1\rangle)/\sqrt{2}$, where 
\begin{eqnarray}\label{eq:01}
\tilde{\Phi}\equiv \sum\limits_{j=1}^{N}\tilde{\phi}_j =\sum\limits_{j=1}^{N}\arccos{(2\sqrt{\epsilon}|a_j\langle P_j\rangle|)}.
\end{eqnarray}
Finally, we perform a projective measurement on the QM qubit to estimate the phase following the standard Kitaev phase estimation procedure~\cite{Kitaev1995}. To obtain a good estimate of $\tilde{\Phi}$, the whole encoding process and measurement process is repeated $M_q$ times.
\par The complication arise from the fact, that since $|\tilde{\Psi}_0\rangle$ is an equal superposition of  $|\phi^{\pm}\rangle$ and $\cos{(\cdot)}$ is an even function, we need to project onto one of the eigenstates $|\phi^{+}\rangle$ or $|\phi^{-}\rangle$   prior to encoding to control the sing of the phase. Note that both eigenstates work since the sign can be controlled by using either $(\tilde{U})_{P_j}(\epsilon)$ or $(\tilde{U})^{\dagger}_{P_j}(\epsilon)$, depending on the eigenstate projection. This is necessary since the end goal is to encode a given linear combination of Pauli strings in the phase of the memory qubit through sequential application of the encoding step with different Pauli strings, $P_j$.

To perform the eigenstate projection, a small series of partial projections (Kitaev's quantum phase estimation (KQPE) steps \cite{Kitaev1995}) are performed prior to the encoding to efficiently resolve between $|\phi^{+}\rangle$ and $|\phi^{-}\rangle$. This can be achieved using an auxiliary qubit as the control qubit~\cite{Wang2019} instead of the memory qubit. It suffices to do enough measurements to accurately determine the sign of the encoded phase in the auxiliary qubit while the actual phase is not necessary to estimate. Consequently, this can be done efficiently with the total number of QPE steps, $n_{QPE}$, scaling logarithmic with the total number of Pauli strings to be encoded assuming that the mean value of each Pauli string is bounded away from zero.
For details, we refer to the Appendices~\ref{app_2}-\ref{app_4}. Importantly, the eigenstate projection is performed using an additional auxiliary qubit and not the memory qubit. This is what allows us to reduce the interactions between the memory qubit and the many-qubit device compared to the CPS method.

\par The steps of our method are the following (see Fig.~\ref{fig:4}):
\begin{itemize}
    \item First, we roughly estimate the mean values of every Pauli string $ \langle\Psi_0|P_j| \Psi_0\rangle$, $j=\overline{1,N}$ by a small series of projective measurements like it is done in the QEE method. This step provides information about the sign and magnitude (bounded away from zero) of the mean values  of the Pauli strings. For more details see Appendix~\ref{app_2}.
\item Second, for every $P_j$ (that is bounded away from zero), we do $n_{QPE}$ of KQPE steps on $|\tilde{\Psi}_0\rangle$, projecting on either $|\phi^{+}\rangle$ and $|\phi^{-}\rangle$.
 If we project on $|\phi^{+}\rangle$, we encode the mean value the Pauli string in the phase of the memory qubit using  $c(\tilde{U})_{P_j}(\epsilon)$ while for $|\phi^{-}\rangle$, we use $c(\tilde{U})^{\dagger}_{P_j}(\epsilon)$.
  For more details see Appendix~\ref{app_3}.
\end{itemize} 
\begin{figure}[ht!]  
\centering 
\includegraphics[width=0.5\linewidth]{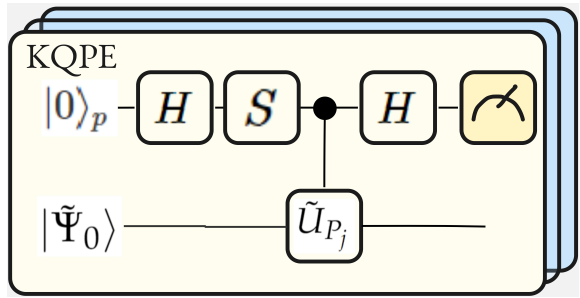}  
\caption{ Quantum circuit proceeding $n_{QPE}$ steps of the Kitaev's QPE  on $|\tilde{\Psi}_0\rangle$ to resolve between $|\phi^{+}\rangle$ and $|\phi^{-}\rangle$. Here $S\in \{I,R_z(\pi/2)\}$, $H$ is a Hadamard gate and $\ket{0}_p$ is a processing ancillary qubit.
}\label{fig:1}  
\end{figure}
For $\epsilon	\ll\,1$, the Taylor expansion is used to rewrite Eq.~\eqref{eq:01}. To lowest order in $\epsilon$, we get:
\begin{equation}
    O\approx\frac{N\pi-2\tilde{\Phi}}{4\sqrt{\epsilon}}.
    \label{26_2}
\end{equation}
The error due to higher order terms in $\epsilon$ can, however, be corrected using additional information from the rough estimates of the mean values used to estimate the signs and magnitudes of the Pauli strings. In Appendices \ref{app_4}-\ref{app_5}, the correction procedure is described in detail and the final variance of the estimate $\hat{O}$ is deduced assuming that $M_c^{cor}$ projective measurements are used for the rough estimates. 
 The variance of the corrected estimate scales as $\sigma^2{(\hat{O}^{TCPS})}\sim N^{\frac{4}{3}}T^{-1}$, where $T$ is the total amount of controlled unitaries used in the TCPS method.  This scaling results as a trade-off between the inaccuracy of the estimate due to the Taylor expansion and the added noise from the correction procedure.
\par Note that in our method, the phase estimation is limited to the modulus of $2\pi$. To address the issue of phase wrapping, we adopt the \textit{sampling} approach introduced in \cite{Higgins2009,Kimmel2015}. Instead of encoding $\phi$ with a fixed $\epsilon$, we employ multiple orders of sampling, denoted as $\tilde{\epsilon}_l\equiv 2^l \sqrt{\epsilon_0}$, where $l=1,2,\dots,d_L$, to gradually zoom in on the value of $\phi$. By appropriately selecting $d_L$ and the total number of measurement repetitions, we achieve the same variance rate. The initial parameter $\epsilon_0$ is chosen in such a way that $\tilde{\Phi}(\epsilon_0)<2\pi$ holds. Further detailed analyses and discussions can be found in Appendix~\ref{app_7}.

\begin{table}[h] 
\caption{Comparison of resources for QEE, CPS and TCPS methods; $t_{prep}$ -  state preparation time; $\eta$ - estimation variance.\label{tab_1}}
\newcolumntype{C}{>{\centering\arraybackslash}X}
\begin{tabularx}{\textwidth}{c|C||C|C||C|}
\hline 
\textbf{Method}	& \textbf{Number of  state preparations}	& \textbf{Qubits}&\textbf{ Coherence times}&\textbf{NISQ to QM interaction}\\
\hline 
QEE 		&$\frac{N^2}{\eta}$ & Processing qubits of  NISQ			& $t_{prep}$&-\\
\hline 
TCPS	&$\!\!\!(\frac{N^{\frac{4}{3}}}{\eta}) O( \log{(\frac{N}{\sqrt{\eta}})})$& Processing qubits of  NISQ			&$t_{prep}O( \log{(\frac{N}{\sqrt{\eta}})})$& $N$\\
&& Memory qubit       &   $N t_{prep}O( \log{(\frac{N}{\sqrt{\eta}})})$ &\\
\hline 
CPS	&$\!\!\!\!(\frac{N}{\eta})O\left(\frac{\log{(\frac{N}{\sqrt{\eta}})}}{\log{\log{(\frac{N}{\sqrt{\eta}})}}}\right)$& Processing qubits of  NISQ			& $t_{prep}O( \log{(\frac{N}{\sqrt{\eta}})})$&  $N O\left(\frac{\log{(\frac{N}{\sqrt{\eta}})}}{\log{\log{(\frac{N}{\sqrt{\eta}})}}}\right)$  \\
&& Memory qubit       &    $\!\!\!\!Nt_{prep}O\left(\frac{\log{(\frac{N}{\sqrt{\eta}})}}{\log{\log{(\frac{N}{\sqrt{\eta}})}}}\right)$&\\
\hline 
\end{tabularx}
\end{table}
We performed a comprehensive comparison of resources between the TCPS, the QEE and CPS, as detailed in Table~\ref{tab_1}. The number of state preparation circuits required and the coherence time of the necessary processing ancillas are compared. To implement $\tilde{U}_{P_j}(\epsilon)$ in the TCPS technique, the controlled sign flip operator $\Pi_0$ and two state preparation circuits $\tilde{V}$ and $\tilde{V}^{\dagger}$ are needed. The controlled $\Pi_0$ can be implemented using a series of two-qubit CNOT gates, which scales linearly with the number of qubits in the NISQ device~\cite{Wang2019,PhysRevA.93.022311} and we therefore view this as having the same cost as the state preparation circuit. 
\par  In terms of time consumption, the state preparation step $t_{prep}$ is considered the most time-consuming, which is also used to quantify the time of the encoding operations.  In the QEE method, the coherence time of the NISQ device is  $t_{prep}$ as each qubit is measured after each state preparation. On the other hand, the TCPS method necessitates a modest increase in the coherence time of the NISQ device. This increase scales logarithmically with the number of Pauli strings and the target variance $\eta$. Additionally, the TCPS method relies on a long coherence time memory qubit, whose coherence time is linearly dependent on the number of Pauli strings $N$. In cases where the available resources cannot provide the required coherence time, the total Pauli sum must be divided into sub-sums and encoded in QM separately.  
\par In return, the TCPS method provides a much better estimate of $\langle O \rangle$ than the QEE method for a fixed number of state preparations. 
Comparing it with the variance of QEE method in the case of equal amount of resources used in both methods, we get
\begin{align}
    \frac{\sigma^2{(\hat{O}^{TCPS})}}{\sigma^2{(\hat{O}^{QEE})}}&\sim\frac{1}{N^{\frac{2}{3}}}.
    \label{1238_1}
\end{align}
which shows that the TCPS method achieves a better scaling of the variance in the number of Pauli strings compared to the scaling of the QEE approach. 
\par As we mentioned in \cite{https://doi.org/10.48550/arxiv.2212.07710} both the  QEE and CPS methods  are subject to the accumulation of operational errors, which poses a limitation on the accuracy of the final estimate of the observable.  The presence of gate errors significantly affects the reliability of both approaches, ultimately impacting the quality of the observables estimation.
However, in contrast to the CPS, the TCPS method has less interactions of the NISQ qubits with the QM.
In TCPS method the encoding of every Pauli string in the QM is done by only one unitary gate. That means, that the amount of operations between NISQ qubits and the QM is $N$ that is significantly less then in the CPS method. This fact makes TCPS method more robust to the QM corruption due to the interactions, however 
in terms of robustness to gate
errors the TCPS and the CPS are similar.

\section{Summary}
We  proposed an alternative variation of our \emph{Coherent Pauli Summation} method  recently provided in \cite{https://doi.org/10.48550/arxiv.2212.07710} called the \emph{Taylor Coherent Pauli Summation} method to estimate the expectation values of multi-qubit  observables.  The method uses phase-kick back techniques and Taylor decomposition to encode information from a multi-qubit processor into a single qubit quantum memory. In this way the  accumulation of the shot noise 
 that arise in the conventional QEE approach is avoided. The variance of the TCPS method outperforms the QEE approach as $\sim 1/N^{2/3}$. 
For observables with a large Pauli string decomposition our method gives a significant advantage, especially if there are many non-commuting Pauli sets. If the Pauli stings can be efficiently sorted in commuting sets 
 they could  be measured in parallel using the QEE method while the TCPS method in the current form in not parallelizable. Thus, our method can be combined with known QEE approaches and applied in cases where sorting into commuting sets is hard (in general it is NP hard) or simply not possible.
 \par In comparison to the CPS, TCPS approach exhibits slightly higher variance but is more robust to the logical errors of the QM. We have demonstrated that TCPS demands $N$ controlled unitary interactions of the NISQ qubits with the QM, making it more experimentally feasible on NISQ devices then CPS.

\section*{Funding} L.M. was partly supported by the Netherlands Organisation for Scientific Research (NWO/OCW), as part of the Quantum Software Consortium program (project number 024.003.037 / 3368).  This research work was partly supported by the Roadmap for the Development of Quantum Technologies in Russian Federation, contract No. 868-1.3-15/15-2021.

\appendix
\section[\appendixname~\thesection]{Hadamard Test}\label{app1}
\numberwithin{equation}{section}
\renewcommand{\theequation}{\thesection\arabic{equation}}
Following the line in \cite{Wang2019}, one of the ways to efficiently  collapse the state $\ket{\tilde{\Psi}_0}$ into one of the eigenstates $\ket{\phi^{\pm}}$ is to run the Hadamard test (Kitaev's QPE) circuit~\cite{Kitaev1995} shown in Fig.~\ref{fig:1}. We don't need to do the full QPE, but as it is shown in Appendix \ref{app_3}, just a small amount of rounds, so the state wont be ruined, but we will extract the information about the $\sign{(\phi)}$ to resolve between two eigenstates with a good confidence rate.  
\par 
The output state of the Hadamard  circuit with $S=I$ can be written as follows
\begin{equation}
    \frac{1}{2}((1+e^{ i\phi})|0\rangle_p+(1-e^{
    i\phi})|1\rangle_p)|
\tilde{\Psi}_0\rangle.
    \label{1502}
\end{equation}
The probability of measuring $|0\rangle$ and $|1\rangle$  are:
\begin{equation}
    P(X=0|\phi)=\frac{1}{2}(1+\cos{(\phi)}),\quad P(X=1|\phi)=\frac{1}{2}(1-\cos{(\phi)}),
    \label{1503}
\end{equation}
that gives a precise estimates of the modulus of the phase for a sufficient amount of iterations. However, \textit{cosine} does not allow us to distinguish between $\phi$ and $-\phi$. Hence, the same Hadamard test is used, but with $S=R_z(\pi/2)$, namely:
\begin{equation}
    \frac{1}{2}((1+ie^{ i\phi})|0\rangle_p+(1-ie^{ i\phi})|1\rangle_p)|\tilde{\Psi}_0\rangle.
    \label{1501}
\end{equation}
The probabilities of measuring $|0\rangle$ and $|1\rangle$  are:
\begin{equation}
    P(Y=0|\phi)=\frac{1}{2}(1-\sin{(\phi)}),\quad P(Y=1|\phi)=\frac{1}{2}(1+\sin{(\phi)}),
    \label{1503_1}
\end{equation}
respectively.
Repeated measurements of these quantum circuits allow to approximate $P(X=0|\phi)$, $P(Y=0|\phi)$ from which the \textit{sine} and \textit{cosine} values and then the $\phi$ itself  can be estimated. 
We denote the estimates of $\sin{(\phi)}$ and $\cos{(\phi)}$ by $\hat{s}$ and $\hat{c}$, respectively.
The \textit{tangent} function is more robust to error then the inverse \textit{sine} and \textit{cosine}  functions (see~\cite{mohammadbagherpoor2019} for details).  We write the estimate as follows:

\begin{equation}
    \hat{\phi}=\arctan{\left(\frac{\hat{s}}{\hat{c}}\right)}.
    \label{1626}
\end{equation}
Since we have two outcomes ($0$ and $1$), we get a Bernoulli distributed independently identically distributed (i.i.d.) sample with probability of success $P(0|\phi)$.
The probability terms are estimated by frequency $\hat{\nu}$ to accuracy:

\begin{equation}
    |\hat{\nu}-P(0|\phi)|< \epsilon_0.
    \label{1258}
\end{equation}
It is straightforward to verify that

\begin{equation}
    |\hat{s}-\sin{(\phi)}|\leq 2\epsilon_0,\quad |\hat{c}-\cos{(\phi)}|\leq 2\epsilon_0,
    \label{1258_1}
\end{equation}
hold.
We want to estimate our angle with high variance, namely

\begin{equation}
    |\phi-\arctan{\left(\frac{\hat{s}}{\hat{c}}\right)}|\leq \epsilon_{tan}.
    \label{1627}
\end{equation}
To find the connection between $\epsilon_{tan}$ and $\epsilon_0$, we consider the case $|\hat{s}-\sin{(\phi)}|=2\epsilon_{0}$, $|\hat{c}-\cos{(\phi)}|=2\epsilon_{0}$ when $\phi=0$. Substituting it in Eq.~\eqref{1627}, we get the following bound:

\begin{equation}
    \epsilon_{0}\leq \frac{\tan{(\epsilon_{tan})}}{2(1\mp \tan{(\epsilon_{tan})})}.
    \label{1628}
\end{equation}
Following~\cite{Kitaev1995} , where $\epsilon_{\tan}=1/16$, holds, we can conclude that $\epsilon_0<1/2(1-1/\sqrt{2})$, holds.
To estimate the amount of measurements we need to guaranty Eq.~\eqref{1258}, we use Chernoff's inequality (see for example~\cite{Berg2020}):

\begin{equation}
    1-2\exp{\left(-2n_0\epsilon_0^2\right)}\leq P\left(|\hat{\nu}_{0}-P(0|\phi)|< \epsilon_0\right).
    \label{1531_25}
\end{equation}
In many applications, it is important to know what the sample size must be in order that, with probability at least $(1- \eta_0)$, one could assert that the estimate  differ from the corresponding value by an amount less than $\epsilon_0$.
In other words, beginning with what value $n_0$, does the following inequality $2\exp{\left(-2n_0\epsilon_0^2\right)}\leq \eta_0$,  $\eta_0\in[0,1]$, holds? One can deduce that the amount of measurements that guarantee this is equal to:

\begin{equation}
    n_0\geq \frac{\log{(2/\eta_0)}}{2\epsilon_0^2}.
    \label{1450}
\end{equation}
Taking into account that each Hadamard test has to be done twice and $m$ angle estimations are done to get enough statistical samples, we get in total $2m$ estimations. Hence, in total we do
\begin{equation}
   M_{K}\geq\frac{m\log{(2/\eta_0)}}{\epsilon_0^2}.
   \label{1615_1}
\end{equation}
\par One can see, that due to the \textit{sine} and \textit{cosine}  functions the method performs bad in the boundary points of $0$ and $1$. That is why we need to do a preliminary check that every $\langle P_j\rangle$ is in the interval bounded away from  $0$ and $1$ done by a small amount of projective measurements. 

\section[\appendixname~\thesection]{On Selecting the Amount of Projective Measurements for the Boundary Condition  Check-up}\label{app_2}
It is known that QPE performs best when the phase is bounded away from $0$ and $2\pi$. Thus, we have to be sure that every   $\langle P\rangle\in \mathcal{I}=[0+\delta,1-\delta]$, where $\delta>0$. That is why we roughly estimate the mean values of every Pauli string $ \langle P_j\rangle$, $j=1,\dots,N$ by a small series of projective measurements like it is done in the QEE method. 
Let $\hat{\langle P\rangle}$ be an estimate of $\langle P\rangle$, using $n_1$ samples. For the independent random variables bounded by the interval $[0, 1]$ the Hoeffding's inequality
\begin{equation}
    P(|\langle P\rangle-\hat{\langle P\rangle}|\geq g_1)\leq 2\exp{(-2n_1g_1^2)},
    \label{1531}
\end{equation}
holds,  where $g_1>0$.
The parameters $g_1$  and $n_1$ can be selected according to our needs of the estimation variance. Reversing Eq.~\eqref{1531}, we get:
\begin{equation}
    1-2\exp{(-2n_1g_1^2)}\leq P(|\langle P\rangle-\hat{\langle P\rangle}|< g_1).
    \label{1531_22}
\end{equation}
Finally, we deduce that
\begin{equation}\label{1637}
    2\exp{(-2n_1g_1^2)}\leq \eta_1,
\end{equation}
holds, where $\eta_1\in[0,1]$ is the desired probability of being inside of $\langle P\rangle\in \mathcal{I}$. The total amount of projective measurements we need for $N$ Paulis is $M_1=Nn_1$.

\section[\appendixname~\thesection]{On $\sign{(\phi)}$ Estimation}\label{app_3}

In the previous section we selected $n_1$ in such a way that with $1-\eta_1$ probability $\langle P\rangle$ is in $\mathcal{I}$. 
Using the part of Kitaev's QPE circuit whose action is given by Eq.~\eqref{1501}, we find that the probability of measuring $|0\rangle$ and $|1\rangle$ is given by  Eq.~\eqref{1503_1}.
If $\phi>0$ the probability $P(Y=0|\phi)<1/2$, holds. We will use this fact to find out the $\sign{(\phi)}$. 

Since $\phi=2\arccos{(|\langle P\rangle|)}$, holds,  we can write:
\begin{align}
    P(Y=0|\phi)=\frac{1}{2}(1-\sin{\phi})\in [&p_{0min},p_{0max}],\quad \text{where}\nonumber\\
    p_{0min}=\frac{1}{2}(1-\sin{(2\arccos{(|1-\delta|)})}),\quad &p_{0max}=\frac{1}{2}(1-\sin{(2\arccos{(|\delta|)})}), \quad \text{if} \quad \delta<1/2,
    \label{1503_1_1}
\end{align}
and
\begin{align}
    P(Y=1|\phi)=\frac{1}{2}(1+\sin{\phi})&\in [p_{1min},p_{1max}],\quad \text{where}\nonumber\\
    p_{1min}=\frac{1}{2}(1+\sin{(2\arccos{(|\delta|)})}),\quad &p_{1max}=\frac{1}{2}(1+\sin{(2\arccos{(|1-\delta|)})}), \quad \text{if} \quad \delta<1/2.
    \label{1503_1_2}
\end{align}
For example, if we select $\delta=0.2$, then the probabilities are defined on the distant intervals $P(Y=0|\phi)\in [0.02,0.3]$, $P(Y=1|\phi)\in [0.7,0.98]$. Thus, for the reasonable selection of $\delta$ the probabilities are spaced apart from $1/2$.

Since our goal is to find $P(Y=0|\phi)\lessgtr 1/2$, we use the Hoeffding's inequality with the following conditions:
\begin{equation}
    1-2\exp{\left(-2g_3^2(\delta)n_{QPE}\right)}\leq P\left(|\hat{\nu}_{Y=0}-P(Y=0|\phi)|< g_3(\delta)\right),
    \label{1531_24}
\end{equation}
where $g_3(\delta)\equiv \frac{1}{2}-p_{0max}(\delta)$ and $\hat{\nu}_{Y=0}$ is a frequency of the event $Y=0$. Beginning with what value $n_{QPE}$, we would like that
\begin{equation}
    2\exp{\left(-2g_3^2(\delta)n_{QPE}\right)}\leq \eta_3,\quad \eta_3\in[0,1],
\end{equation}
holds? One can deduce that it is bounded as follows
\begin{equation}\label{1638}
    n_{QPE}\geq \frac{\log{(2/\eta_3)}}{2g_3^2(\delta)}.
\end{equation}
If all $\langle P_j\rangle\in\mathcal{I}$ we run the algorithm for $N$ phases and get the total amount of QPE rounds  bounded by
\begin{equation}
    M_{QPE}\geq\frac{N\log{(2/\eta_{3})}}{2g_3^2(\delta)}.
    \label{1615_1046}
\end{equation}
Finally, setting $M_{QPE}$ rounds,  with probability $1-\eta_3$ we will estimate $P(Y=0|\phi)$ by $\hat{\nu}_{Y=0}$ with enough variance to determine $P(Y=0|\phi)\lessgtr \frac{1}{2}$.

\section[\appendixname~\thesection]{Variance Estimation}\label{app_4}

\begin{figure}[ht!]  
\centering 
\includegraphics[width=0.5\linewidth]{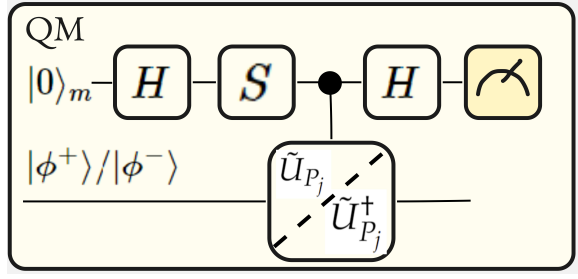} 
\caption{The phase-kick back process, encoding the mean values of the Pauli strings in the phase of the state stored in the memory qubit. Here $S$ is identity or $R_z(\pi/2)$ gate.
}\label{fig:2}
\end{figure}

After the sign estimation done by $n_{QPE}$ rounds of QPE we do the phase encoding in the memory qubit $|\psi\rangle_{m}=|+\rangle_m$, applying c$\tilde{U}_{P}$ or c$\tilde{U}_{P}^{\dagger}$ as it is depicted in Fig.~\ref{fig:2} with $S=I$. 
This process is repeated $N$ times until all the phases connected with the Pauli strings are encoded in the long coherence time QM. 
The accumulated state encoded in $|\psi\rangle_{m}$ is 
\begin{equation}
   |\tilde{\Phi}\rangle= \frac{|0\rangle_m+e^{ i\tilde{\Phi}}|1\rangle_m}{\sqrt{2}}|\phi_1\rangle|\phi_2\rangle\dots |\phi_N\rangle,\quad \tilde{\Phi}=\sum\limits_{i=1}^N\arccos{(2\sqrt{\epsilon}|a_j\langle P_j\rangle|)}.
    \label{1058}
\end{equation}
Then we apply the Hadamart gate
\begin{equation}
   H\otimes I|\tilde{\Phi}\rangle= \frac{(1+e^{ i\tilde{\Phi}})|0\rangle_m+(1-e^{ i\tilde{\Phi}})|1\rangle_m}{2}|\phi_1\rangle|\phi_2\rangle\dots |\phi_N\rangle.
    \label{1058_2}
\end{equation}
Similarly to the previous sections, the probabilities to measure $|0\rangle$ and $|1\rangle$ are
\begin{equation}
    P(X=0|\tilde{\Phi})=\frac{1}{2}(1+\cos{(\tilde{\Phi})}),\quad P(X=1|\tilde{\Phi})=\frac{1}{2}(1-\cos{(\tilde{\Phi})}).
    \label{1235}
\end{equation}
If use the encoding with $S=R_z(\pi/2)$,  we get
\begin{equation}
   |\tilde{\Phi}\rangle= \frac{|0\rangle_m+ie^{ i\tilde{\Phi}}|1\rangle_m}{\sqrt{2}}|\phi_1\rangle|\phi_2\rangle\dots |\phi_N\rangle.
    \label{1324}
\end{equation}
Then we apply the Hadamart gate on the latter state , we get the following result
\begin{equation}
   H\otimes I|\tilde{\Phi}\rangle= \frac{(1+ie^{ i\tilde{\Phi}})|0\rangle+(1-ie^{ i\tilde{\Phi}})|1\rangle}{2}|\phi_1\rangle|\phi_2\rangle\dots |\phi_N\rangle
    \label{1058_4}
\end{equation}
and the probabilities to measure $|0\rangle_m$ and $|1\rangle_m$ are
\begin{equation}
    P(Y=0|\tilde{\Phi})=\frac{1}{2}(1+\sin{(\tilde{\Phi})}),\quad P(Y=1|\tilde{\Phi})=\frac{1}{2}(1-\sin{(\tilde{\Phi})}).
    \label{1235_2}
\end{equation}
 Since we can't prepare the reversed  preparation circuit of the state $|\tilde{\Phi}\rangle$, we can't use QPE to estimate the phase directly. To estimate
$P(X=0|\tilde{\Phi})$, $P(Y=0|\tilde{\Phi})$ and then find the total accumulated phase we need to repeat all the procedure of encoding $M_q$ times, every time doing the projective measurement on the state encoded in the QM.
Then, using the probability  estimates, we estimate \textit{sine} and \textit{cosine} that will determine the estimate of $\tilde{\Phi}$.
To this end, we introduce the notation 
\begin{equation}
    \hat{Q}\equiv \frac{1-2\hat{P}(Y=0| \tilde{\Phi})}{2\hat{P}(X=0| \tilde{\Phi})-1}=\hat{\tan}{(\tilde{\Phi})},
    \label{21}
\end{equation}
and the variance of the estimate of ${\Phi}$ is the following
\begin{equation}
\sigma^2\hat{\tilde{\Phi}}\approx\left(\frac{\partial \tilde{\Phi}}{\partial Q}\right)^2\sigma^2\hat{Q}=\left(\frac{1}{1+Q^2}\right)^2\sigma^2\hat{Q}.
    \label{22_1}
\end{equation}
The variance of $\hat{Q}$ can be written as follows
\begin{equation}
    \sigma^2\hat{Q}\approx\sigma^2\hat{P}(X=0| \tilde{\Phi})\left(\frac{2(1-2\hat{P}(Y=0| \tilde{\Phi}))}{(2\hat{P}(X=0| \tilde{\Phi})-1)^2}\right)^2+\sigma^2\hat{P}(Y=0| \tilde{\Phi})\left(\frac{2}{2\hat{P}(X=0| \tilde{\Phi})-1}\right)^2.
    \label{22}
\end{equation}
Since we have a Bernoulli distributed random variables, the variances are
\begin{equation}
    \sigma^2\hat{P}(X=1| \tilde{\Phi})=\frac{\hat{P}(X=0| \tilde{\Phi})\hat{P}(X=1|\tilde{\Phi})}{M_q},\quad \sigma^2\hat{P}(Y=1| \tilde{\Phi})=\frac{\hat{P}(Y=0| \tilde{\Phi})\hat{P}(Y=1| \tilde{\Phi})}{M_q},
    \label{24_2}
\end{equation}
where $M_q$ is the amount of repreparations of $|\tilde{\Phi}\rangle$.
Finally, the variance \eqref{22_1} can be written as follows:
\begin{equation}
    \sigma^2\hat{\tilde{\Phi}}\approx\frac{3 + \cos{(8\tilde{\Phi})} }{4 M_q}\sim \frac{1}{ M_q}.
    \label{22_3}
\end{equation}
Then the variance of the estimate of $O$ is the following
\begin{equation}
 \sigma^2\hat{O}\approx\left(\frac{\partial O}{\partial\phi}\right)^2\sigma^2{\hat{\tilde{\Phi}}}\sim
    \frac{1}{ M_q\epsilon}.
    \label{26_3_1226}
\end{equation}
\section{Error Correction}\label{app_5}
\par Since we used the Taylor expansion to get rid of \textit{arc-cosine} function in \eqref{1058}, we inserted an error by neglecting the $O$-big term.Hence, we use the QEE strategy  to estimate the correction terms. The corrected value is the following
\begin{equation}
    \hat{O}_{TCPS} =\hat{O}+\frac{1}{2\sqrt{\epsilon}}\sum\limits_{n=1}^{\infty}\frac{(2n)!(2\sqrt{\epsilon})^{2n+1}}{4^n(n!)^2(2n+1)}\hat{P}^{2n+1}_{cor},
    \label{27_1336}
\end{equation}
where 
\begin{equation}
\hat{P}^{2n+1}_{cor}=\sum\limits_{i=1}^{N}|a_i^{2n+1}\langle \hat{P}_{i}\rangle^{2n+1}|,
     \label{27_1336_2}
\end{equation}
is the sum of the Pauli strings estimates  done by the projective measurements (like it is done in QEE method).
The variance of the correction term can be straightforwardly written as
follows
\begin{equation}
    \sigma^2{{P^{2n+1}_{cor}}}=\frac{(2n+1)}{n_c^{cor}}\sum\limits_{j=1}^{N}(a_j^{2n+1}\langle P_j\rangle^{2n})^2(1-\langle P_j\rangle^2),
    \label{27_2}
\end{equation}
where $ n_c^{cor}$ is the amount of the projective measurements done for every Pauli string. Then 
$M_c^{cor}=N n_c^{cor}$ is the amount of the projective measurement done to estimate $N$ correction terms. Hence, the variance of the estimate of $O$ is the following
\begin{equation}
    \sigma^2{\hat{O}_{TCPS} }=\sigma^2{\hat{O}}+\frac{1}{n_c^{cor}}\sum\limits_{j=1}^{N}\sum\limits_{n=1}^{\infty}
    \frac{((2n)!)^22^{4n}\sqrt{\epsilon}^{4n}}{2^{4n}(n!)^4}a_j^{2(2n+1)}\langle P_j\rangle^{4n}(1-\langle P_j\rangle^2).
    \label{27_4}
\end{equation}
The series in the right hand side  is of the  type:
\begin{equation}
    \sum\limits_{n=1}^{\infty}\frac{((2n)!)^2(2\sqrt{\epsilon}a_j\langle P_j\rangle)^{4n}}{2^{4n}(n!)^4}=\frac{2}{\pi}K\left(16\sqrt{\epsilon}^4a_j^4\langle P_j\rangle^4\right)-1\quad \mbox{if}\quad \sqrt{\epsilon}^2a_j^2\langle P_j\rangle^2<1/4,
    \label{50_8}
\end{equation}
where 
\begin{equation}
    K(t)=\int_{0}^{\pi/2}\frac{d\theta}{\sqrt{1-t\sin^2{\theta}}},
    \label{27_5}
\end{equation}
is the complete elliptic integral of the first kind\footnote{One must be careful with the notation when using these functions, because various reputable references and software packages use different conventions in the definitions of the elliptic functions. On Wikipedia $K(k)$ is used, $k^2=t$, holds.}. Finally, we can rewrite Eq.~\eqref{27_4} as follows
\begin{equation}
    \sigma^2\hat{O}_{TCPS}  =\frac{1}{\epsilon M_q}+\frac{1}{n_c^{cor}}\sum\limits_{j=1}^{N}\Big[
    \frac{2}{\pi}K\left(16\epsilon^2 a_j^4\langle P_j\rangle^4\right)-1\Big]a_j^2(1-\langle P_j\rangle^2),
    \label{27_6}
\end{equation}
if
\begin{equation}
    \epsilon<\left(4a_j^2\langle P_j\rangle^2\right)^{-1},
    \label{27_7}
\end{equation}
holds for all $j=1,\dots, N$. Let us use the following asymptotic expression:
\begin{equation}
    K\left(t\right)\approx\frac{\pi}{2}+\frac{\pi}{8}\frac{t}{1-t}-\frac{\pi}{16}\frac{t^2}{1-t}.
    \label{27_8}
\end{equation}
This approximation has a relative variance better than $3\times 10^{-4}$ for $t<1/4$ ($k<1/2$). Keeping only the first two terms is correct to $0.01$ variance for $t<1/4$ ($k<1/2$).
\par We can conclude that one can use \eqref{27_8} if $t\equiv \epsilon^2 a_j^4\langle P_j\rangle^4<1/4$. Since Eq.~\eqref{27_7}, holds, it is always true. Using this assumption, we can rewrite Eq.~\eqref{27_6} as follows:

\begin{equation}
    \sigma^2{\hat{O}_{TCPS} }\sim \frac{1}{\epsilon M_q}+\frac{\epsilon^2}{n_c^{cor}}\sum\limits_{j=1}^{N}
    a_j^6\langle P_j\rangle^4(1-\langle P_j\rangle^2).
    \label{27_99}
\end{equation}
We select $\epsilon<<1$. Taking the derivative on $\epsilon$ and equating it to zero, we get the optimal $\epsilon$ that minimizes the variance \eqref{27_99}:
\begin{equation}
    \epsilon=\sqrt[3]{\frac{n_{c}^{cor}}{M_{q}\sum\limits_{j=1}^{N} a_j^6\langle P_j\rangle^4(1-\langle P_j\rangle^2)}}\sim
    \left(\frac{n_c^{cor}}{M_q}\right)^{\frac{1}{3}} \frac{1}{N^{\frac{1}{3}}}= \left(\frac{M_c^{cor}}{M_q}\right)^{\frac{1}{3}} \frac{1}{N^{\frac{2}{3}}}.
    \label{27_11_3}
\end{equation}
\par  The standard deviation of our estimate should be bigger or equal to the error value that can be done on the sign estimation and QPE steps, namely $\sum\limits_{j=1}^{N}2a_j\langle P_j\rangle((1-\eta_1)\eta_3+\eta_1/2)$. 
The parameters $\eta_1$ and $\eta_3$ are defined in \eqref{1637} and \eqref{1638} as
\begin{eqnarray}\label{1140}
\eta_1=\frac{2}{e^{2n_1g_1^2}}
,\quad \eta_3=\frac{2}{e^{2n_{QPE}g_3^2(\delta)}}.
\end{eqnarray}
Let us assume $\eta_1$ being the same rate as $\eta_3$.
Since $\eta_1<<1$, we can write
\begin{eqnarray}\label{27_10_1413_1}
(1-\eta_1)\eta_3+\eta_1/2\sim \eta_3+\eta_1/2\sim e^{-2n_{QPE}g_3^2(\delta)}.
\end{eqnarray}
If we assume that all Paulis are equal with equal weights, then we can write the condition 
\begin{eqnarray}\label{1804}
\sigma^2{\hat{O}_{TCPS}} \geq 36a^2\langle P\rangle^2 N^2 e^{-4n_{QPE}g_3^2(\delta)} .
\end{eqnarray}
 Finally  if we define the target variance $\sigma^2{\hat{O}_{TCPS}}=\eta$, we can conclude
\begin{eqnarray}\label{1805}
n_{QPE}=O\left(\log{\left(\frac{N}{\sqrt{\eta}}\right)}\right)
\end{eqnarray}
that gives the scaling of QPE steps required for our method.
\section[\appendixname~\thesection]{Comparison of TCPS Method and QEE Method}\label{app_6}

In this section we  compare the TCPS with the QEE. 
In QEE every Pauli string is measured independently and then all the results are summed up. Hence, the variance of the estimate of $O$ is the following:

\begin{align}
    \sigma^2{\hat{O}_{QEE}}&\equiv \sum\limits_{i=1}^{N}\frac{a_i^2\sigma^2{\hat{P}_i}}{n_c}=\frac{N}{M_c}\sum\limits_{i=1}^{N}a_i^2\left(1-\langle\Psi_0| P_i|\Psi_0\rangle^2\right)\sim\frac{N^2}{M_c},
    \label{eq:1049}
\end{align}
where $M_c=N n_c$ is the amount of projective  measurements  done in QEE.
Then
\begin{eqnarray}\label{27_11}
\frac{\sigma^2{\hat{O}_{TCPS}}}{\sigma^2{\hat{O}_{QEE}}}\sim\frac{\frac{1}{M_q^{\frac{2}{3}}(n_c^{cor})^{\frac{1}{3}}}\left(\sum\limits_{j=1}^{N}
   a_j^6\langle P_j\rangle^4(1-\langle P_j\rangle^2)\right)^{\frac{1}{3}}}{\frac{1}{n_{c}} \sum\limits_{j=1}^{N}a_j^2(1-\langle P_j\rangle^2)}.
\end{eqnarray}

Let us  assume that the amount of resources needed for both methods are equal. For TCPS we have the total amount of resources $M_T=M_{c}^{cor}+M_{q}(1+M_{QPE})$. Hence we select $M_c=M_{T}$ and, substituting \eqref{27_11_3} in \eqref{27_11}, we get
\begin{eqnarray}\label{27_11_4}
\frac{\sigma^2{\hat{O}_{TCPS}}}{\sigma^2{\hat{O}_{QEE}}}\sim\frac{M_{q}(1+M_{QPE})+M_{c}^{cor}}{N^{\frac{2}{3}}M_{q}^{\frac{2}{3}}(M_{c}^{cor})^{\frac{1}{3}}} 
\frac{\left(\sum\limits_{j=1}^{N}a_j^6\langle P_j\rangle^4 (1-\langle P_j\rangle^2)\right)^{\frac{1}{3}}}{\sum\limits_{j=1}^{N}a_j^2(1-\langle P_j\rangle^2)}.
\end{eqnarray}
Since we are interested in the rate, we assume the case when all $P_j$ and $a_j$ are equal for all $j$. Then the latter ratio will reduce to
\begin{eqnarray}\label{27_11_5}
\frac{\sigma^2{\hat{O}_{TCPS}}}{\sigma^2{\hat{O}_{QEE}}}\sim\frac{M_{q}(1+M_{QPE})+M_{c}^{cor}}{N^{\frac{4}{3}}M_{q}^{\frac{2}{3}}(M_{c}^{cor})^{\frac{1}{3}}}
\frac{\langle P\rangle^{\frac{4}{3}}}{(1-\langle P\rangle^2)^{\frac{2}{3}}}.
\end{eqnarray}
Since \eqref{1805}, holds, and selecting $M_c^{cor}=M_q(1+M_{QPE})$ and $M_c^{cor}=N m_c^{cor}$, we can conclude 
\begin{eqnarray}\label{27_11_1350}
\frac{\sigma^2{\hat{O}_{TCPS}}}{\sigma^2{\hat{O}_{QEE}}}\sim \frac{1}{N^{\frac{2}{3}}}.
\end{eqnarray}
\section[\appendixname~\thesection]{Control of Phase Wrapping} \label{app_7}
The state \eqref{1058} encoded in the memory qubit contains the phase
\begin{eqnarray}\label{1522_1}
\hat{\Phi}=\sum\limits_{j=1}^{N}\arccos{\left(2\sqrt{\epsilon}|a_j\langle P_j\rangle|\right)}\mod{2\pi},
\end{eqnarray}
accumulated by the $N$ rounds of encoding. However, the phase in our method is estimated only up to modulus $2\pi$.
With the growth of $N$ the phase $\hat{\Phi}\notin [-\pi,\pi)$ and  one has no information about how many times the phase wraps around $2\pi$.
To overcome this problem we can use method 
introduced in \cite{Higgins2009}, and improved in \cite{Kimmel2015} for the purpose of gate
calibration.
We select $\tilde{\epsilon}_0\equiv 2\sqrt{\epsilon_0}$ and introduce the notation
\begin{eqnarray}\label{9011}
\phi_0=\tilde{\epsilon}_0P_{sum},\quad P_{sum}\equiv\sum\limits_{j=0}^{N} |a_j\langle P_j\rangle|,
\end{eqnarray}
 such that $\phi_0<2\pi$, holds.
Given
a targeted variance $\eta > 0$, and numbers $\alpha,\gamma\in  Z^{+}$, the algorithm outputting
an estimate $\hat{P}_{sum}$ as an estimate for ${P}_{sum}$ proceeds as follows. We 
 fix $d_L=[\log_2{1/\eta}]$ and for all $l=1,2,3,\dots, d_L$ obtain estimates $\hat{\phi}_l$ of $\phi_l= 2^l \tilde{\epsilon}_0 P_{sum} \mod{2\pi}$  from $M_l=\alpha+\gamma(d_L+1-l)$ repetitions of Hadamard test circuit with $S=I$ and $S=R$. For $l=1$ we set  $\hat{P}^{(1)}_{sum}\equiv \frac{\hat{\phi}_0}{\tilde{\epsilon}_0}$.
  For all other $l$ parameters we set $\hat{P}^{(l)}_{sum}$ to be the (unique) number in  $[\hat{P}^{(l-1)}_{sum}-\frac{\pi}{2^{l}},\hat{P}^{(l-1)}_{sum}+\frac{\pi}{2^{l}})$ such that 
\begin{eqnarray}\label{3002}2^{l}\tilde{\epsilon}_0\hat{P}^{(l)}_{sum}\equiv \hat{\phi}_{l-1},\end{eqnarray} holds.
 Finally, after $l$ steps, the $\hat{P}_{sum}=\hat{P}^{(d_L)}_{sum}$ returns an estimate of ${P}_{sum}$.
\par In \cite{Kimmel2015} it is shown that  choosing  $\alpha>2$ and $\gamma>0$, the variance $\eta$
 of the final estimate scales as $\sim cM^{-1}$, where where  $c$ is a constant  and $M$ is a total cost.
\par Then, the total amount of repetitions of the Hadamard test is $M_q=2\sum\limits_{l=0}^{d_L} M_l$ and the variance of the estimate is of the rate
\begin{eqnarray}\label{3006}\sigma^2(\hat{O}_{TCPS})\sim\frac{1}{2^{2(d_L+1)}\epsilon M_q}+\frac{\epsilon ^2 N^2}{M_c^{cor}}.\end{eqnarray}
  The optimal $\epsilon$ selection that minimizes the variance is
\begin{equation}
    \epsilon=\left(\frac{M_c^{cor}}{2^{2(d_l+1)}M_q }\right)^{\frac{1}{3}} \frac{1}{N^{\frac{2}{3}}}.
    \label{3008}
\end{equation}
Then  the variance \eqref{3006} can be written as follows
  \begin{eqnarray}\label{4006}\sigma^2(\hat{O}_{TCPS})\sim\frac{N^{\frac{2}{3}}}{M_q^{\frac{2}{3}}(M_c^{cor})^{\frac{1}{3}}2^{\frac{4}{3}d_L}}.\end{eqnarray}
The total amount of the  state preparations used in the Taylor based method is $T=(1+ n_{QPE})NM_q+M_c^{cor}$ and then the final expression is the following
 \begin{eqnarray}\label{3009}\sigma^2(\hat{O}_{TCPS})\sim
 \frac{N^{\frac{4}{3}}n_3^{\frac{2}{3}}}{T 2^{\frac{4}{3}d_L}},\end{eqnarray}
where we still get $1/N^{\frac{2}{3}}$ overhead in variance comparable to QEE method.
\section{Implementing TCPS}
As we mentioned, the TCPS algorithm  requires multi-qubit controlled gates/complex connectivity's to perform. To begin with, the error rate of the controlled reflection gate $\Pi_0$  and the controlled Pauli operations required for $\tilde{U}_{P_j}$ must be smaller then the errors of the state preparation gate $\tilde{V}$.
\par The multi-qubit controlled gates can be achieved for example on Rydberg or trapped ion devices. In ion trap systems, the controlled reflection operator can be implemented using various techniques to achieve quantum control. One common approach involves applying laser-induced vibrational sideband transitions, where the internal states of the ions are coupled to their motional modes. By carefully engineering the laser pulses and controlling the ion trap parameters, it is possible to achieve controlled reflections \cite{gulde2003implementation,monroe2013scaling}.
Moreover, the Rydberg atom array platform provides a promising avenue for the implementation of the controlled reflection operator and controlled Pauli operations, with a favorable scaling characteristic that grows linearly with the system size \cite{molmer2011efficient,isenhower2011multibit}. 
\par Next, TCPS requares a long coherence QM. The ion-qubits have an ultra-long coherence time \cite{langer2005long,wang2017single} and near perfect qubit state initialization and detection. In \cite{wang2021single} a single ion  $^{171}Yb^{+}$ QM with the coherence time more then one hour (5500 s) is implemented. The experimental demonstration shows its applicability on NISQ devices and robustness to the various noises. For example, $99.99\%$ detection fidelities are demonstrated in \cite{myerson2008high} and the shortest detection time $\sim 11\mu s$ is provided with $99.93\%$ fidelity in \cite{crain2019high}. 
For the single qubit gates, it has been shown
that the duration of the gates approach scales  picoseconds and the fidelities are much higher than the typical error correction requirements  with both microwaves and laser beams \cite{cai2023entangling}.
In \cite{PhysRevLett.127.130505} the  $35\mu s$ duration  two-qubit gates with a fidelity of $99.94\%$ are realised on optical ions of $^{40}Ca^{+}$. In \cite{schafer2018fast} the  $1.6\mu s$ duration  two-qubit gates with a fidelity of $99.8\%$ are realised on hyperfine ions of $^{43}Ca^{+}$.
\par Unfortunately, the ion traps  have strong interaction with environmental and control noises that is the source of decoherence of qubit states and gate operations. For example, in a fully connected ion trap system where every ion can interact with any other ion, the number of elementary operations required to implement a two-qubit gate typically scales quadratically with the number of ions. This means that the gate time or the number of physical operations required for the gate execution increases as the square of the system size. 
  \par That is why at that moment atomic systems involving highly excited Rydberg states are an attractive system for our method.
  In \cite{li2016quantum} it is suggested to employ Rydberg levels for interactions and ground levels for storage to achieve both fast quantum operations and long-lived memory ($\sim 70-80\mu s$).
  In \cite{PRXQuantum.3.010344} an architecture for quantum
computing, which takes small, high-fidelity, local
quantum processors, places them inside an optical cavity,
and quickly connects them using, heralded single-photon
transfers is proposed. This idea was applied to  Rydberg atoms and
multiple chains of trapped ions. This architecture looks promising for our protocol since contains good Rydberg atom based gates connected to long-coherence quantum ion memory.

\bibliographystyle{unsrt} 
\bibliography{fer_12}

\begin{thebibliography}{10}

\bibitem{Golub2000}
G.H. Golub and H.A. {van der Vorst}.
\newblock Eigenvalue computation in the 20th century.
\newblock {\em J. Comput. Appl. Math}, 123(1):35--65, 2000.
\newblock Numerical Analysis 2000. Vol. III: Linear Algebra.

\bibitem{Kitaev1995}
A.~Yu. Kitaev.
\newblock Quantum measurements and the abelian stabilizer problem.
\newblock {\em Conferance}, 1995.

\bibitem{Kimmel2015}
S.~Kimmel, G.H. Low, and T.J. Yoder.
\newblock Robust calibration of a universal single-qubit gate set via robust
  phase estimation.
\newblock {\em Phys. Rev. A}, 92:062315, 2015.

\bibitem{Berg2020}
E.~van~den Berg.
\newblock Iterative quantum phase estimation with optimized sample complexity.
\newblock In {\em 2020 IEEE International Conference on Quantum Computing and
  Engineering (QCE)}, pages 1--10, 2020.

\bibitem{mohammadbagherpoor2019}
H.~Mohammadbagherpoor, Y.H. Oh, P.~Dreher, A.~Singh, X.~Yu, and A.J. Rindos.
\newblock An improved implementation approach for quantum phase estimation on
  quantum computers.
\newblock In {\em 2019 IEEE International Conference on Rebooting Computing
  (ICRC)}, pages 1--9. IEEE, 2019.

\bibitem{Wiebe2016}
N.~Wiebe and C.~Granade.
\newblock Efficient bayesian phase estimation.
\newblock {\em Phys. Rev. Lett.}, 117:010503, Jun 2016.

\bibitem{O_Brien2019}
T.E. O'Brien, B.~Tarasinski, and B.M. Terhal.
\newblock Quantum phase estimation of multiple eigenvalues for small-scale
  (noisy) experiments.
\newblock {\em New J. Phys}, 21(2):023022, feb 2019.

\bibitem{Peruzzo2014}
A.~Peruzzo, J.~McClean, P.~Shadbolt, M.-H. Yung, X.-Q. Zhou, P.J. Love,
  A.~Aspuru-Guzik, and J.L. O'Brien.
\newblock A variational eigenvalue solver on a photonic quantum processor.
\newblock {\em Nat. Commun}, 5, 2014.

\bibitem{jurcevic2021}
P.~Jurcevic, A.~Javadi-Abhari, L.S. Bishop, I.~Lauer, D.F. Bogorin, M.~Brink,
  L.~Capelluto, O.~G{\"u}nl{\"u}k, T.~Itoko, N.~Kanazawa, et~al.
\newblock Demonstration of quantum volume 64 on a superconducting quantum
  computing system.
\newblock {\em Quantum Sci. Technol.}, 6(2):025020, 2021.

\bibitem{brown2016}
K.R. Brown, J.~Kim, and C.~Monroe.
\newblock Co-designing a scalable quantum computer with trapped atomic ions.
\newblock {\em npj Quantum Inf.}, 2(1):1--10, 2016.

\bibitem{schafer2018}
V.M. Sch{\"a}fer, C.J. Ballance, K.~Thirumalai, L.J. Stephenson, T.G. Ballance,
  A.M. Steane, and D.M. Lucas.
\newblock Fast quantum logic gates with trapped-ion qubits.
\newblock {\em Nature}, 555(7694):75--78, 2018.

\bibitem{Wang2019}
D.~Wang, O.~Higgott, and S.~Brierley.
\newblock Accelerated variational quantum eigensolver.
\newblock {\em Phys. Rev. Lett.}, 122:140504, Apr 2019.

\bibitem{Hamamura2020}
I.~Hamamura and T.~Imamichi.
\newblock Efficient evaluation of quantum observables using entangled
  measurements.
\newblock {\em npj Quantum Inf.}, 6:2056--6387, 2020.

\bibitem{Crawford2021}
O.~Crawford, B.~Straaten, D.~Wang, T.~Parks, E.~Campbell, and S.~Brierley.
\newblock Efficient quantum measurement of {P}auli operators in the presence of
  finite sampling error.
\newblock {\em {Quantum}}, 5:385, January 2021.

\bibitem{McClean2016}
J.~R. McClean, J.~Romero, R.~Babbush, and A.~Aspuru-Guzik.
\newblock The theory of variational hybrid quantum-classical algorithms.
\newblock {\em New J. Phys}, 18(2):023023, feb 2016.

\bibitem{https://doi.org/10.48550/arxiv.2212.07710}
L.A. Markovich, A.~Almasi, S.~Zeytinoglu, and J.~Borregaard.
\newblock Quantum memory assisted observable estimation, 2022.

\bibitem{PhysRevLett.118.010501}
G.H. Low and I.L. Chuang.
\newblock Optimal hamiltonian simulation by quantum signal processing.
\newblock {\em Phys. Rev. Lett.}, 118:010501, Jan 2017.

\bibitem{langer2005long}
C.~Langer, R.~Ozeri, J.D. Jost, J.~Chiaverini, B.~DeMarco, A.~Ben-Kish, R.B.
  Blakestad, J.~Britton, D.B. Hume, W.M. Itano, et~al.
\newblock Long-lived qubit memory using atomic ions.
\newblock {\em Phys. rev. lett.}, 95(6):060502, 2005.

\bibitem{wang2017single}
Y.~Wang, M.~Um, J.~Zhang, S.~An, M.~Lyu, J.-N. Zhang, L.-M. Duan, D.~Yum, and
  K.~Kim.
\newblock Single-qubit quantum memory exceeding ten-minute coherence time.
\newblock {\em Nature Phot.}, 11(10):646--650, 2017.

\bibitem{li2016quantum}
L.~Li and A.~Kuzmich.
\newblock Quantum memory with strong and controllable rydberg-level
  interactions.
\newblock {\em Nat. Comm.}, 7(1):13618, 2016.

\bibitem{RevModPhys.87.307}
B.M. Terhal.
\newblock Quantum error correction for quantum memories.
\newblock {\em Rev. Mod. Phys.}, 87:307--346, Apr 2015.

\bibitem{doi:10.1080/09500340.2012.737937}
J.R. Wootton.
\newblock Quantum memories and error correction.
\newblock {\em Journal of Modern Optics}, 59(20):1717--1738, 2012.

\bibitem{McClean2014}
J.~R. McClean, R.~Babbush, P.~J. Love, and A.~Aspuru-Guzik.
\newblock Exploiting locality in quantum computation for quantum chemistry.
\newblock {\em J. Phys. Chem. Lett}, 5(24):4368--4380, 2014.
\newblock PMID: 26273989.

\bibitem{jordan1993paulische}
Pascual Jordan and Eugene~Paul Wigner.
\newblock {\"U}ber das paulische {\"a}quivalenzverbot.
\newblock In {\em The Collected Works of Eugene Paul Wigner}, pages 109--129.
  Springer, 1993.

\bibitem{https://doi.org/10.48550/arxiv.1706.03637}
Panagiotis~Kl. Barkoutsos, Nikolaj Moll, Peter W.~J. Staar, Peter Mueller,
  Andreas Fuhrer, Stefan Filipp, Matthias Troyer, and Ivano Tavernelli.
\newblock Fermionic hamiltonians for quantum simulations: a general reduction
  scheme, 2017.

\bibitem{Knill2007}
E.~Knill, G.~Ortiz, and R.~D. Somma.
\newblock Optimal quantum measurements of expectation values of observables.
\newblock {\em Phys. Rev. A}, 75:012328, Jan 2007.

\bibitem{Higgins2009}
B.L. Higgins, D.W. Berry, S.D. Bartlett, M.W. Mitchell, H.M. Wiseman, and G.J.
  Pryde.
\newblock Demonstrating heisenberg-limited unambiguous phase estimation without
  adaptive measurements.
\newblock {\em New J. Phys}, 11(7):073023, jul 2009.

\bibitem{PhysRevA.93.022311}
D.~Maslov.
\newblock Advantages of using relative-phase toffoli gates with an application
  to multiple control toffoli optimization.
\newblock {\em Phys. Rev. A}, 93:022311, Feb 2016.

\bibitem{gulde2003implementation}
S.~Gulde, M.~Riebe, G.P.T. Lancaster, C.~Becher, J.~Eschner, H.~H{\"a}ffner,
  F.~Schmidt-Kaler, I.L. Chuang, and R.~Blatt.
\newblock Implementation of the deutsch--jozsa algorithm on an ion-trap quantum
  computer.
\newblock {\em Nature}, 421(6918):48--50, 2003.

\bibitem{monroe2013scaling}
C.~Monroe and J.~Kim.
\newblock Scaling the ion trap quantum processor.
\newblock {\em Science}, 339(6124):1164--1169, 2013.

\bibitem{molmer2011efficient}
K.~M{\o}lmer, L.~Isenhower, and M.~Saffman.
\newblock Efficient grover search with rydberg blockade.
\newblock {\em J. Phys. B: At. Mol. Opt. Phys}, 44(18):184016, 2011.

\bibitem{isenhower2011multibit}
L.~Isenhower, M.~Saffman, and K.~M{\o}lmer.
\newblock Multibit c k not quantum gates via rydberg blockade.
\newblock {\em Quantum Inf. Process.}, 10:755--770, 2011.

\bibitem{wang2021single}
P.~Wang, C.-Y. Luan, M.~Qiao, M.~Um, J.~Zhang, Y.~Wang, X.~Yuan, M.~Gu,
  J.~Zhang, and K.~Kim.
\newblock Single ion qubit with estimated coherence time exceeding one hour.
\newblock {\em Nature communications}, 12(1):233, 2021.

\bibitem{myerson2008high}
A.H. Myerson, S.C. Szwer, D.J.and~Webster, D.T.C. Allcock, M.J. Curtis,
  G.~Imreh, J.A. Sherman, D.N. Stacey, A.M. Steane, and D.M. Lucas.
\newblock High-fidelity readout of trapped-ion qubits.
\newblock {\em Phys. Rev. Lett.}, 100(20):200502, 2008.

\bibitem{crain2019high}
S.~Crain, C.~Cahall, G.~Vrijsen, E.E. Wollman, M.D. Shaw, V.B. Verma, S.W. Nam,
  and J.~Kim.
\newblock High-speed low-crosstalk detection of a 171yb+ qubit using
  superconducting nanowire single photon detectors.
\newblock {\em Comm. Phys.}, 2(1):97, 2019.

\bibitem{cai2023entangling}
Z.~Cai, C.-Y. Luan, L.~Ou, H.~Tu, Z.~Yin, J.-N. Zhang, and K.~Kim.
\newblock Entangling gates for trapped-ion quantum computation and quantum
  simulation.
\newblock {\em Journal of the Korean Physical Society}, pages 1--19, 2023.

\bibitem{PhysRevLett.127.130505}
C.R. Clark, H.N. Tinkey, B.C. Sawyer, A.M. Meier, K.~A. Burkhardt, C.M. Seck,
  C.M. Shappert, N.D. Guise, C.E. Volin, S.D. Fallek, H.T. Hayden, W.G.
  Rellergert, and K.R. Brown.
\newblock High-fidelity bell-state preparation with $^{40}{\mathrm{ca}}^{+}$
  optical qubits.
\newblock {\em Phys. Rev. Lett.}, 127:130505, Sep 2021.

\bibitem{schafer2018fast}
V.M. Sch{\"a}fer, C.J. Ballance, K.~Thirumalai, L.J. Stephenson, T.G. Ballance,
  A.M. Steane, and D.M. Lucas.
\newblock Fast quantum logic gates with trapped-ion qubits.
\newblock {\em Nature}, 555(7694):75--78, 2018.

\bibitem{PRXQuantum.3.010344}
J.~Ramette, J.~Sinclair, Z.~Vendeiro, A.~Rudelis, M.~Cetina, and
  V.~Vuleti\ifmmode~\acute{c}\else \'{c}\fi{}.
\newblock Any-to-any connected cavity-mediated architecture for quantum
  computing with trapped ions or rydberg arrays.
\newblock {\em PRX Quantum}, 3:010344, Mar 2022.

\end{thebibliography}
\end{document}